\documentclass[a4paper, fleqn, usenatbib, useAMS]{tex/mnras}

\usepackage{newtxtext, newtxmath}
\usepackage[T1]{fontenc}
\usepackage{ae, aecompl}

\usepackage{comment}
\usepackage{graphicx}
\graphicspath{{figs/}}
\DeclareGraphicsExtensions{.pdf, .ps, .eps, .jpg}
\usepackage[multidot]{grffile}

\usepackage{xspace}
\usepackage[usenames, dvipsnames]{xcolor}

\usepackage{listings}
\definecolor{codegreen}{rgb}{0,0.6,0}
\definecolor{codegray}{rgb}{0.5,0.5,0.5}
\definecolor{codepurple}{rgb}{0.58,0,0.82}
\definecolor{backcolour}{rgb}{0.95,0.95,0.92}
\lstdefinestyle{mystyle}{
    backgroundcolor=\color{backcolour},   
    commentstyle=\color{codegreen},
    keywordstyle=\bf\color{magenta},
    numberstyle=\tiny\color{codegray},
    stringstyle=\color{codepurple},
    basicstyle=\ttfamily\footnotesize,
    breakatwhitespace=false,         
    breaklines=true,                 
    captionpos=b,                    
    keepspaces=true,                 
    numbers=left,                    
    numbersep=5pt,                  
    showspaces=false,                
    showstringspaces=false,
    showtabs=false,                  
    tabsize=2
}
 
\definecolor{grassgreen}{RGB}{0, 186, 0}

\newcommand{\etal}{et\thinspace al.\xspace}
\newcommand{\hii}{H\thinspace\textsc{ii}\xspace}
\newcommand{\starlight}{\textsc{starlight}\xspace}

\newcommand{\Ha}{\ifmmode \text{H}\alpha \else H$\alpha$\fi\xspace}
\newcommand{\Hb}{\ifmmode \text{H}\beta \else H$\beta$\fi\xspace}
\newcommand{\neiii}{\ifmmode [\text{Ne}\,\textsc{iii}] \else [Ne~{\scshape iii}]\fi\xspace}
\newcommand{\Neiii}{\ifmmode [\text{Ne}\,\textsc{iii}]\lambda 3869 \else [Ne~{\scshape iii}]$\lambda 3869$\fi\xspace}
\newcommand{\oi}{\ifmmode [\text{O}\,\textsc{i}] \else [O~{\scshape i}]\fi\xspace}
\newcommand{\Oi}{\ifmmode [\text{O}\,\textsc{i}]\lambda 5577 \else [O~{\scshape i}]$\lambda 5577$\fi\xspace}
\newcommand{\oii}{\ifmmode [\text{O}\,\textsc{ii}] \else [O~{\scshape ii}]\fi\xspace}
\newcommand{\Oii}{\ifmmode [\text{O}\,\textsc{ii}]\lambda 3726 + \lambda 3729 \else [O~{\scshape ii}]$\lambda 3726 + \lambda 3729$\fi\xspace}
\newcommand{\Oiiit}{\ifmmode [\text{O}\,\textsc{iii}]\lambda 4363 \else [O~{\scshape iii}]$\lambda 4363$\fi\xspace}
\newcommand{\heii}{\ifmmode \text{He}\,\textsc{ii} \else He~{\scshape ii}\fi\xspace}
\newcommand{\Heii}{\ifmmode \text{He}\,\textsc{ii}\lambda 4686 \else He~{\scshape ii}$\lambda 4686$\fi\xspace}
\newcommand{\ariv}{\ifmmode [\text{Ar}\,\textsc{iv}] \else [Ar~{\scshape iv}]\fi\xspace}
\newcommand{\Ariv}{\ifmmode [\text{Ar}\,\textsc{iv}]\lambda 4711 + \lambda 4740 \else [Ar~{\scshape iv}]$\lambda 4711 + \lambda 4740$\fi\xspace}
\newcommand{\Niit}{\ifmmode [\text{N}\,\textsc{ii}]\lambda 5755 \else [N~{\scshape ii}]$\lambda 5755$\fi\xspace}
\newcommand{\hei}{\ifmmode \text{He}\,\textsc{i} \else He~{\scshape i}\fi\xspace}
\newcommand{\Hei}{\ifmmode \text{He}\,\textsc{i}\lambda 5876 \else He~{\scshape i}$\lambda 5876$\fi\xspace}
\newcommand{\nii}{\ifmmode [\text{N}\,\textsc{ii}] \else [N~{\scshape ii}]\fi\xspace}
\newcommand{\Nii}{\ifmmode [\text{N}\,\textsc{ii}]\lambda 6584 \else [N~{\scshape ii}]$\lambda 6584$\fi\xspace}
\newcommand{\oiii}{\ifmmode [\text{O}\,\textsc{iii}] \else [O~{\scshape iii}]\fi\xspace}
\newcommand{\Oiii}{\ifmmode [\text{O}\,\textsc{iii}]\lambda 5007 \else [O~{\scshape iii}]$\lambda 5007$\fi\xspace}
\newcommand{\oiiis}{\ifmmode [\text{O}\,\textsc{iii}]_\text{S} \else [O~{\scshape iii}]$_\text{S}$\fi\xspace}
\newcommand{\sii}{\ifmmode [\text{S}\,\textsc{ii}] \else [S~{\scshape ii}]\fi\xspace}
\newcommand{\Sii}{\ifmmode [\text{S}\,\textsc{ii}]\lambda 6716 \else [S~{\scshape ii}]$\lambda 6716$\fi\xspace}
\newcommand{\ariii}{\ifmmode [\text{Ar}\,\textsc{iii}] \else [Ar~{\scshape iii}]\fi\xspace}
\newcommand{\Ariii}{\ifmmode [\text{Ar}\,\textsc{iii}]\lambda 7135 \else [Ar~{\scshape iii}]$\lambda 7135$\fi\xspace}
\newcommand{\siii}{\ifmmode [\text{S}\,\textsc{iii}] \else [S~{\scshape iii}]\fi\xspace}
\newcommand{\Siii}{\ifmmode [\text{S}\,\textsc{iii}]\lambda 9069 \else [S~{\scshape iii}]$\lambda 9069$\fi\xspace}

\newcommand{\rOii}{ \ifmmode [\text{O}\,\textsc{ii} ]\lambda 3726/3729 \else [O~{\scshape  ii}]$\lambda 3726/3729$\fi\xspace}
\newcommand{\rOiii}{\ifmmode [\text{O}\,\textsc{iii}]\lambda 4363/5007 \else [O~{\scshape iii}]$\lambda 4363/5007$\fi\xspace}
\newcommand{\rAriv}{\ifmmode [\text{Ar}\,\textsc{iv}]\lambda 4740/4711 \else [Ar~{\scshape iv}]$\lambda 4740/4711$\fi\xspace}
\newcommand{\rNii}{ \ifmmode [\text{N}\,\textsc{ii} ]\lambda 5755/6584 \else [N~{\scshape  ii}]$\lambda 5755/6584$\fi\xspace}
\newcommand{\rSiii}{\ifmmode [\text{S}\,\textsc{iii}]\lambda 6312/9532 \else [S~{\scshape iii}]$\lambda 6312/9532$\fi\xspace}
\newcommand{\rSii}{ \ifmmode [\text{S}\,\textsc{ii} ]\lambda 6731/6717 \else [S~{\scshape  ii}]$\lambda 6731/6717$\fi\xspace}

\newcommand{\WHa}{\ifmmode W_{\Ha} \else $W_{\Ha}$\fi\xspace}
\newcommand{\Dn}{\ifmmode \mathrm{D}_n 4000 \else $\mathrm{D}_n 4000$\fi\xspace}
\title[DIG effects on O/H estimates in SF galaxies] % <= 45 chars
      {Diffuse ionized gas and its effects on nebular metallicity estimates of star-forming galaxies}

\author[N.\ Vale Asari \etal]
       {N. Vale Asari,$^{1,2}$\thanks{email: natalia@astro.ufsc.br}\thanks{Royal Society--Newton Advanced Fellowship}
        G. S. Couto,$^{3,1}$
        R. Cid Fernandes,$^1$
        G. Stasi\'nska,$^4$\
        \and
        A. L. de Amorim,$^1$
        D. Ruschel-Dutra,$^1$
        A. Werle,$^{5,1}$
        T. Z. Florido$^1$
        \\
        $^{1}$Departamento de F\'{\i}sica--CFM, Universidade Federal de Santa Catarina, C.P.\ 476, 88040-900, Florian\'opolis, SC, Brazil \\
        $^{2}$School of Physics and Astronomy, University of St Andrews, North Haugh, St Andrews KY16 9SS, UK\\
        $^{3}$Centro de Astronom\'{\i}a (CITEVA), Universidad de Antofagasta, Avenida Angamos 601, Antofagasta, Chile \\
        $^{4}$LUTH, CNRS, Observatoire de Paris, PSL University, France \\
        $^{5}$Instituto de Astronomia, Geof\'{\i}sica e Ci\^{e}ncias Atmosf\'{e}ricas, Universidade de S\~{a}o Paulo, R. do Mat\~{a}o 1226, 05508-090 S\~{a}o Paulo, Brazil \\
       }

\date{Accepted \dots. Received \today; in original form \dots}

\pubyear{2018}

\begin{document}

\label{firstpage}
\pagerange{\pageref{firstpage}--\pageref{lastpage}}

\maketitle

%**********************************************************************%
%                                                                      %
%                               Abstract                               %
%                                                                      %
%**********************************************************************%
% https://academic.oup.com/mnras/pages/General_Instructions
% <= 250 words for Main Journal papers or 200 words for Letters
\begin{abstract}

We investigate the impact of the diffuse ionized gas (DIG) on abundance determinations in star-forming (SF) galaxies.
The DIG is characterised using the \Ha equivalent width (\WHa).
From a set of 1\,409 SF galaxies from the Mapping Nearby Galaxies at APO (MaNGA) survey, we calculate the fractional contribution of the DIG to several emission lines using high-$S/N$ data from SF spaxels (instead of using noisy emission-lines in DIG-dominated spaxels).
Our method is applicable to spectra with observed \WHa $\gtrsim 10 \text{~\AA}$ (which are not dominated by DIG emission). Since the DIG contribution depends on galactocentric distance, we provide DIG-correction formulae for both entire galaxies and single aperture spectra.
Applying those to a sample of $\,> 90\,000$ SF galaxies from the Sloan Digital Sky Survey, we find the following.
(1) The effect of the DIG on strong-line abundances depends on the index used. It is negligible for the (\oiii/\Hb)/(\nii/\Ha) index, but reaches $\sim 0.1$ dex at the high-metallicity end for \nii/\Ha.
(2) This result is based on the $\sim$kpc MaNGA resolution, so the real effect of the DIG is likely greater. 
(3) We revisit the mass--metallicity--star formation rate (SFR) relation by correcting for the DIG contribution in \emph{both} abundances and SFR. The effect of DIG removal is more prominent at higher stellar masses. Using the \nii/\Ha index, O/H \emph{increases} with SFR at high stellar mass, contrary to previous claims.

\end{abstract}


%\natalia{\ojo Natalia: Remove this \\clearpage before submitting -- it is here to avoid compilation errors.}
%\clearpage

%**********************************************************************%
%                                                                      %
%                           Introduction                               %
%                                                                      %
%**********************************************************************%
\section{Introduction}
\label{sec:Intro}

The diffuse ionized gas (DIG) is a warm ($\sim 10^4\,$K), low density ($\sim 10^{-1}\,$cm$^{-3}$) gas phase found in the interstellar medium of galaxies \citep[e.g.][]{Haffner.etal.2009a}. While early studies of this faint line-emitting component (usually referred to as the warm ionized medium in the context of our Galaxy) focused on the surroundings of the Milky Way disk \citep{Hoyle.Ellis.1963a, Reynolds.1984a}, the DIG has also been observed in different galaxy structures such as extraplanar regions in edge-on disks \citep{Boettcher.etal.2016a, Levy.etal.2019a}, between \hii regions and interarms in face-on spirals \citep{Greenawalt.etal.1998a, Kreckel.etal.2016a}, and also in spheroids (bulges and ellipticals, \citealp{Gomes.etal.2016c}). Tentative explanations for the ionization of the diffuse gas include leakage of Lyman continuum photons from star-forming regions \citep[e.g.][]{Weilbacher.etal.2018a}; photoionization by hot low-mass evolved stars (HOLMES; \citealp{Stasinska.etal.2008a, FloresFajardo.etal.2011a}); fast shocks from supernova winds \citep{Dopita.Sutherland.1995a, Allen.etal.2008a}; turbulent dissipation \citep{Binette.etal.2009a}; and heating by cosmic rays or dust grains \citep{Reynolds.Cox.1992a}. However, the properties of the DIG and how it affects physical measurements of galaxies are still debatable.

From narrow-band \Ha imaging, \citet{Zurita.Rozas.Beckman.2000a} and \citet{Oey.etal.2007a} estimated that the DIG contributes to 30--60 per cent of the total \Ha flux in local spiral galaxies. Yet the contribution from this tenuous but ubiquitous component to the emission line fluxes is generally neglected in the analysis of integrated (i.e.\ not spatially resolved) galaxy spectra.  When estimating a galaxy's star formation rate from its \Ha luminosity, for instance, one usually assumes that all the observed \Ha results from photoionization by massive young stars. Similarly, estimates of the nebular oxygen abundance in star-forming (SF) galaxies neglect the contribution of the DIG to lines involved in the calculation of O/H (see \citealp{Sanders.etal.2017b} and references therein). Clearly, it would be useful to quantify to which extent these and other emission-line diagnostics are affected by the DIG.

\citet{ErrozFerrer.etal.2019a}, studying 38 galaxies with high spatial resolution observed with the Multi Unit Spectroscopic Explorer  \citep[MUSE;][]{Bacon.etal.2010a}, found that metallicities in \hii regions and the DIG have consistent radial distributions, but median values in \hii regions are $\sim 0.1$~dex higher than in the DIG. Also making use of MUSE, \citet{Kumari.etal.2019a} selected 24 nearby spiral galaxies to study the DIG metallicity. Defining pairs of \hii and DIG regions that are close enough to have the same metallicity, they measure the metallicity difference of each DIG region relative to its \hii region counterpart when applying the same strong-line calibrations. With this procedure, they propose corrections for the effect of the DIG to several metallicity indicators.  

In this study, we propose a different approach.  We estimate the DIG contribution by comparing the \emph{total} galaxy emission in a given aperture with the emission attributed \emph{only to SF regions} in that same aperture. Our analysis thus does not depend on faint emission lines typical of the DIG, but instead relies on higher signal-to-noise emission lines from either SF or SF+DIG regions.

\defcitealias{Lacerda.etal.2018a}{L18}

It is also important to use a reasonable criterion to define what the DIG is. Several criteria have been considered to date. For example, \citet{Blanc.etal.2019a} define the DIG as zones of high \sii/\Ha (this method is also used by \citealp{ErrozFerrer.etal.2019a}), whereas \citet{Zhang.etal.2017c} define the DIG as zones of low surface brightness in \Ha ($\Sigma_{\Ha}$; a scheme also applied by \citealp{Sanders.etal.2017b}).
A combination of \sii/\Ha and $\Sigma_{\Ha}$ is employed by \citet{Blanc.etal.2009a}, \citet{Kaplan.etal.2016a} and \citet{Poetrodjojo.etal.2019a}, who explore criteria tailored for individual galaxies observed with integral field spectroscopy (IFS).
Using IFS data from the CALIFA survey \citep{Sanchez.etal.2012a, Sanchez.etal.2016a, Husemann.etal.2013a, GarciaBenito.etal.2015a}, \citet[][hereafter L18]{Lacerda.etal.2018a} define the DIG as zones of low \Ha equivalent width (\WHa). As argued in that paper, a definition of the DIG must be related to the total stellar population covered by the aperture. For example, galaxy bulges may have quite a high $\Sigma_{\Ha}$, yet -- in galaxies devoid of active galactic nuclei (AGN) and star formation in their central parts -- this zone is ionized by HOLMES \citep{Gomes.etal.2016c} and clearly corresponds to the DIG. Such zones are not identified as DIG using the $\Sigma_{\Ha}$ criterion, but are correctly identified as DIG using the \WHa criterion (see fig.~1 in \citetalias{Lacerda.etal.2018a}).

Applying this definition to CALIFA galaxies, \citetalias{Lacerda.etal.2018a} found systematic variations in \Ha flux fractions along the Hubble sequence, ranging from a dominant DIG in ellipticals and S0 ($f^{\rm DIG}_{\Ha} > 90$ per cent), to a dominant SF component in Sc and later type spirals ($f^{\rm SF}_{\Ha} > 90$ per cent). Intermediate fractions are obtained for intermediate Hubble types, e.g.\ DIG fractions of $\sim 50$ per cent for Sb galaxies. It is worth recalling that the bimodal distribution of $\WHa$ at the root of this scheme is also seen in integrated galaxy data \citep{CidFernandes.etal.2010a, CidFernandes.etal.2011a}.

In this paper we apply this same DIG-identification framework to an 8 times larger sample from the Mapping Nearby Galaxies at APO \cite[MaNGA,][]{Bundy.etal.2015a} survey. Our central goal is to evaluate the effects of DIG emission upon line luminosities observed in integrated galaxy spectra, and then to estimate how this affects nebular metallicities obtained from strong line methods which are also applied to integrated galaxy spectra. In particular we will examine how this impacts the mass--metallicity relation ($M$--$Z$), more specifically in its version where the star formation rate (SFR) acts as a second parameter, the $M$--$Z$--SFR relation.

It is important to stress from the outset that our ability to correct for DIG effects is limited by the spatial resolution of the data. This limitation should nonetheless be contrasted with the case of integrated spectroscopy, where the lack of spatial information mixes different line emitting regimes into a total of inseparable components. In this context, despite of their limitations, the DIG corrections shown here represent a step forward, even if only first order corrections.

The outline of the paper is as follows. Section \ref{sec:Data} describes the MaNGA data and how it is processed. Section \ref{sec:WHa_MaNGA} presents the distribution of $\WHa$ as a function of radius and host morphology, similarly to what has been done by \citetalias{Lacerda.etal.2018a} for CALIFA. Section \ref{sec:howtocorrect} explains the procedure we propose to correct line luminosities measured from single-aperture spectra for the DIG contribution. Section \ref{sec:metallicity estimates} quantifies the bias in the metallicities obtained from various line indicators when neglecting the effect of the DIG. In Section \ref{sec:MZSFR} we show how contamination by DIG emission affects the relations between mass, metallicity, and SFR for SF galaxies in the seventh data release (DR7, \citealp{Abazajian.etal.2009a}) of the Sloan Digital Sky Survey (SDSS, \citealp{York.etal.2000a}). Section \ref{sec:summary} summarises our main findings.  Throughout this work, we assume a flat $\Lambda$CDM cosmology with $\Omega_0 = 0.3$ and $H_0 = 70\,\text{km}\,\text{s}^{-1}\,\text{Mpc}^{-1}$.

%**********************************************************************%
%                                                                      %
%                               Data                                   %
%                                                                      %
%**********************************************************************%
\section{MaNGA Data}
\label{sec:Data}

MaNGA \citep{Bundy.etal.2015a} is an IFS survey from SDSS IV \citep{Blanton.etal.2017a} that will observe $\sim$10000 galaxies across $\sim$2700 $\mathrm{deg}^2$ of sky up to $z\sim0.03$ from 360 to 1000 nm with resolution $R\sim2000$. The observations are made with integral field units with 19 to 127 fibres of 2 arcsec in diameter, yielding a spatial sampling of 1--2 kpc. In this work, we use the 4\,824 currently available MaNGA galaxy datacubes from SDSS DR15 \citep{Aguado.etal.2019a}.  Redshifts used in this work come from the NASA-Sloan Atlas \citep[NSA, ][]{Blanton.etal.2011a} catalogue catalogue (extracted from the MaNGA \texttt{drpall} table version 2.4.3); 4\,679 datacubes have redshifts measured.

Most of the data used in this work are obtained directly from MaNGA spectra.  We also use morphological classifications from the Galaxy Zoo I project \citep{Lintott.etal.2008a, Lintott.etal.2011a}.  The concentration index CI is taken as the ratio between the 90 and the 50 per cent Petrosian light radius (i.e.\ $R_{90}$ and $R_{50}$) in the $r$-band as obtained from the SDSS DR7 database \citep{Abazajian.etal.2009a}.

%**********************************************************************%

%**********************************************************************%
\subsection{Preprocessing steps and measurements}
\label{sec:MaNGA_datacubes}

We use linearly-sampled reduced datacubes downloaded from \url{https://data.sdss.org/sas/dr15/manga/spectro/redux/v2_4_3/[PLATE]/stack/}.
The cubes were preprocessed according to the following steps:

\begin{enumerate}

\item For quality assurance, we use the \texttt{MASK} extension from the datacubes, which contains flags for each spatial position and wavelength, and for whole spaxels.  We mask spaxels with low fibre coverage or foreground stars, and we also mask bad values from the spectra (flagged as \texttt{DONOTUSE}).
  
\item MaNGA spectra are stored using vacuum wavelengths. Our analysis is based on data in air wavelengths. We convert vacuum to air wavelengths using the IAU standard conversion, given in \cite{Morton.1991a}.

\item The cubes are spatially resampled to $1 \,\mathrm{arcsec} \times 1 \,\mathrm{arcsec}$ pixels. This amounts to a $2 \times 2$ binning of the original $0.5 \,\mathrm{arcsec} \times 0.5 \,\mathrm{arcsec}$ sampling of the MaNGA cubes, which is an oversampling given the $\sim 2.5 \,\mathrm{arcsec}$ FWHM of the PSF.  The fluxes are added together, and the error spectra are added in quadrature.  Due to the way the cube is reconstructed in the MaNGA pipeline, spaxels are spatially correlated.  To take the covariance into account, we multiply the combined errors by the factor $1 + 1.62\ \log\ N$, with $N = 4$, according to the recipe given by \cite{Law.etal.2016a}. Fluxes and errors in binned spaxels with less than 4 components, due to masked data, are scaled assuming the missing data is equal to the average of the binned spaxel.  This is done separately for each wavelength.  If all components are masked, the binned spaxel is masked.

\item After the spatial resampling, we further discard most binned spaxels with $S/N < 3$ around $5635$ \AA, but keeping a few of them to `fill in the image' using a convex hull algorithm, to avoid non-contiguous images.
Next, we discard all binned spaxels where $>50$ per cent of spectral points were masked in the previous step.
Three datacubes (\texttt{plate-ifudsgn} identifications: 8312-6101, 8317-1902 and 8332-1901) have been discarded at this step, as they contained no spaxel with S/N $\geq 3$ around 5635 \AA. Thus only the remaining 4\,676 cubes have gone through the rest of the preprocessing steps described below.

\item We correct for the the Galactic extinction using the \citet*{Cardelli.Clayton.Mathis.1989a} dust law with $R_V = 3.1$.  The color index $E(B - V)$ is extracted from the \texttt{FITS} header.
These values were obtained from \cite*{Schlegel.Finkbeiner.Davis.1998a} dust maps.

\item The spectra are shifted to the rest frame, and resampled to the wavelength range 3\,600--10\,400 with $\Delta\lambda =  1\,\text{\AA}$. 

\item Modelling of stellar continua is done with the inverse spectral synthesis code \starlight \citep{CidFernandes.etal.2005a}. We model the observed spectrum (masking out bad pixels, emission line spectral windows, and sky features) as a sum of stellar populations attenuated by a single dust screen following the \citet{Cardelli.Clayton.Mathis.1989a} dust law with $R_V = 3.1$.

Although the MaNGA pipeline does a fairly good job at removing the sky emission, it may fail in some cases. This is especially visible in O~{\scshape i}$\lambda 5577$ and Na$\lambda 5896$ (sodium from street lamps). These lines are flagged as sky features to \starlight -- data around these lines are kept in the resulting spectra, but are not used in the spectral fit. The emission from OH radicals creates a `forest' of lines in the red part of the spectra, which can also be tricky to be removed correctly. For simplicity, we choose not to fit the spectra in wavelengths above $8900\,\text{\AA}$.

The \starlight fits use the 2016 updated spectral synthesis models by \citet{Bruzual.Charlot.2003a}\footnote{\url{http://www.bruzual.org/~gbruzual/cb07/Updated_version_2016/}}, with a \citet{Chabrier.2003a} initial mass function, MILES stellar library below 7351 \AA\ \citep{SanchezBlazquez.etal.2006a,FalconBarroso.etal.2011a} combined with STELIB stellar library at redder wavelengths \citep{LeBorgne.etal.2003a}, and `Padova (1994)' \citep{Alongi.etal.1993a,Bressan.etal.1993a,Fagotto.etal.1994a,Fagotto.etal.1994b,Girardi.etal.1996a} evolutionary tracks. Theoretical spectra from the Tlusty \citep{Lanz.Hubeny.2003a, Lanz.Hubeny.2003b}, Martins \citep{Martins.etal.2005a}, UVBlue \citep{RodriguezMerino.etal.2005a} and PoWR \citep{Sander.Hamann.Todt.2012a} libraries are downgraded in spectral resolution and used to complement observed stars in the MILES stellar library.  We use a base of 96 composite stellar populations obtained by assuming constant star formation rates in 16 logarithmically-spaced age bins between 1\,Myr and 14\,Gyr and 6 metallicities between $Z = 0.0001$ to $0.05$ (similar to the base used by \citealp{Werle.etal.2019a}, but there they used not-yet-public updated \citeauthor{Bruzual.Charlot.2003a} models).

\item We assemble all data and \starlight products into a fits file for analysis within the \textsc{pycasso} software \citep{CidFernandes.etal.2013a, deAmorim.etal.2017a}.

\item Emission line measurements. See the following section.
  
\end{enumerate}
%**********************************************************************%

%**********************************************************************%
\subsection{Emission line measurements}
\label{sec:MaNGA_elf}

Two codes were used to model emission lines: {\scshape ifscube}\footnote{{\scshape ifscube} is a free Python package for the analysis of integral field spectroscopy data cubes, which includes an emission line fitter (ELF), and is available at \url{https://bitbucket.org/danielrd6/ifscube}.} and {\scshape dobby}\footnote{{\scshape dobby} is a free ELF available at \url{https://bitbucket.org/streeto/pycasso2}.}. Earlier tests with the MaNGA 14th data release cubes showed that measurements with {\scshape dobby} and {\scshape ifscube} are consistent; in the following we thus show details and results for {\scshape dobby}.

Emission line fluxes are measured fitting Gaussian profiles to the residual spectra, which are obtained by subtracting the \starlight modeled continua from the observed spectra.  {\scshape dobby} uses the {\scshape astropy} fitting model \citep{AstropyCollaboration.etal.2013a, AstropyCollaboration.etal.2018a}, which calls the {\scshape scipy} \citep{Scipy.2019a} wrapper for the {\scshape minpack} implementation \citep{Garbow.Hillstrom.More.1980a} of the Levenberg--Marquardt algorithm.

The fits obey a number of a priori `common sense' physical constraints: lines from the same ion must have the same kinematics (same central velocity and intrinsic velocity dispersion)\, we require the $\Ha/\Hb$ line flux ratio to be $\geq 2.6$, and the $\Oiii/\oiii4959$ and $\Nii/\nii6548$ line ratios to be fixed according to the values predicted by atomic physics. Imperfections in the stellar continuum model are accounted for by allowing for a local pseudo-continuum in the fits, modeled as a Legendre polynomial of degree 16.

Besides emission line fluxes ($F$) for a given emission line, we also compute equivalent widths ($W$). The continuum flux density $C$ used to compute $W = F / C$ is evaluated with the standard procedure of calculating the median continua on side-bands and using a linear interpolation to measure the continuum at the line centre. Table \ref{tab:EmLinesContinuumSideBands} shows the line centres and side bands for all emission lines used in this work. We find it more convenient to measure $C$ in the \starlight synthetic spectra, as it is free of noise artefacts.

The total width of the Gaussian profile is calculated by adding the intrinsic and the instrumental widths in quadrature, where the latter is taken from the dispersion spectra in the original MaNGA cubes.  The noise $\sigma_N$ on the continuum under emission lines is taken to be the rms in the detrended residual continua in the two windows blueward and redward of the lines. The amplitude-to-noise ratio ($A/N$) of an emission line is the Gaussian amplitude divided by $\sigma_N$. Following \cite{Rola.Pelat.1994a}, the uncertainty in flux measurements is taken to be $\sigma_F = \sigma_N \sqrt{6 \sigma_\lambda \Delta\lambda}$, where $\sigma_\lambda$ is the Gaussian dispersion in \AA, and $\Delta\lambda = 1$~\AA\ is the spectral sampling. The signal-to-noise ratio of an emission line is thus given by $S/N = F / \sigma_F$.

\begin{table}
    \centering
    \caption{Emission line continuum pass bands.}
    \begin{tabular}{lccc}
        \hline
        Line               & Line centre [\AA] & Blue Cont. [\AA] & Red Cont.[\AA]\\
        \hline
        $\oii\lambda3726$  & $3726.032$ & $3653$--$3713$ & $3741$--$3801$ \\
        $\oii\lambda3729$  & $3728.815$ & $3653$--$3713$ & $3741$--$3801$ \\ 
        \Hb                & $4861.330$ & $4770$--$4830$ & $4890$--$4910$ \\
        $\oiii\lambda4959$ & $4958.911$ & $4915$--$4950$ & $5015$--$5050$ \\
        \Oiii              & $5006.843$ & $4915$--$4950$ & $5015$--$5050$ \\
        $\nii\lambda6548$  & $6548.040$ & $6480$--$6530$ & $6600$--$6670$ \\ 
        \Ha                & $6562.820$ & $6480$--$6530$ & $6600$--$6670$ \\
        \Nii               & $6583.460$ & $6480$--$6530$ & $6600$--$6670$ \\ 
        $\sii\lambda6716$  & $6716.440$ & $6650$--$6700$ & $6745$--$6790$ \\ 
        $\sii\lambda6731$  & $6730.810$ & $6650$--$6700$ & $6745$--$6790$ \\ 
        \hline
    \end{tabular}
    \label{tab:EmLinesContinuumSideBands}
\end{table}

%**********************************************************************%

%**********************************************************************%

\subsection{The MaNGA master sample}
\label{sec:Sample}

From the 4\,676 fully preprocessed datacubes, we have selected only those with no observation or calibration problems (i.e.\ with a data quality flag \texttt{DRP3QUAL} = 0, which means that no bitmask value is set on this flag, as explained in \citealp{Law.etal.2016a}, and whose table 16 is updated online at \url{https://www.sdss.org/dr13/algorithms/bitmasks/#MANGA_DRP3QUAL}).  We also removed repeated observations of the same galaxies, keeping only the observation made with the largest number of fibres, or, in case of a tie, the greatest exposure time. This left us with 4\,131 galaxies.

We have then applied the following cuts to our sample:
\begin{enumerate}
\item After looking at SDSS images for galaxies in our preprocessed sample, we decided to remove very edge-on systems by selecting galaxies with an axial ratio $b/a \ge 0.3$ (as measured by NSA, using the column \texttt{nsa\_elpetro\_ba} from \texttt{drpall}).

\item We then removed objects whose probability of being in a merging system is $> 0.4$ \citep{Darg.etal.2010a}.
Objects for which there is no such information in Galaxy Zoo I are also removed.

\item A small fraction of the remaining galaxies contain too many masked spaxels, due to the removal of foreground objects in the field. To filter out the worst cases we kept galaxies with at least 80 per cent useful spaxels (see Section~\ref{sec:MaNGA_datacubes}) within $1\, R_{50}$ from the centre of the observation (defined as the reference pixel in the data cubes).
\end{enumerate}

Our final master sample of 3\,236 galaxies is comprised of all objects that satisfy those cuts and are tagged either as Primary+ or as Secondary MaNGA targets.  This master sample is used in Section \ref{sec:WHa_MaNGA}, where we validate the DIG identification methodology proposed by \citetalias{Lacerda.etal.2018a}.

%**********************************************************************%

%**********************************************************************%
%                                                                      %
%                            Histograms                                %
%                                                                      %
%**********************************************************************%
\section{The distribution of the equivalent width of \Ha in  MaNGA spaxels}
\label{sec:WHa_MaNGA}

Analysing data from the CALIFA survey \citep{Sanchez.etal.2012a, Sanchez.etal.2016a, Husemann.etal.2013a, GarciaBenito.etal.2015a}, \citetalias{Lacerda.etal.2018a} found a strongly bimodal distribution of $\WHa$ (also noted by \citealp{Belfiore.etal.2016a} using MaNGA data), with peaks at $\sim 1$ and 14 \AA\ and a minimum at $\sim 3$ \AA. This lead them to the following classification scheme:

\begin{itemize}
\item $\WHa < 3$ \AA\ zones trace the HOLMES-powered DIG, or hDIG. Such zones  appear frequently in spheroids (throughout elliptical galaxies and also in bulges), as well as above and below disks. Their energetics, line ratios, and locations are all consistent with a HOLMES origin although a contribution from shocks is not excluded.

\item  $\WHa > 14$ \AA\ zones are associated with star-forming complexes (SFc), typical of disks. SFc trace SF sites in a galaxy, although, because of the 1--2 kpc resolution, they cannot be considered pure SF regions. 

\item $\WHa = 3$--14 \AA\ zones constitute the mixed DIG (mDIG), a buffer category where line emission may be understood as a mixture of hDIG and SFc, but which probably includes other processes as well, such as photon leakage from \hii regions, OB runaways, or shocks.
\end{itemize}

It must be noted that, while the 3 \AA\ limit corresponds to the minimum of the distribution in $\WHa$ among galaxies and is effectively closely linked to the value expected for ionisation by a purely old stellar population \citep[see][]{CidFernandes.etal.2011a}, the 14 \AA\  limit is arbitrary, and serves only to separate regions that contain a good proportion of SF-powered line emission  from regions whose ionization is due to a mixture of processes.

It is also important to mention that, at the $\sim$ kpc resolution of MaNGA (also CALIFA), even regions classified as SFc will contain some diffuse gas emission. This is why we prefer the `SF complex' terminology instead of the more standard `SF region', which semantically suggests a regime where only \hii regions contribute to the line  emission.

The hDIG/mDIG/SFc classification of the nebular regime introduced in \citetalias{Lacerda.etal.2018a} is rooted on the bimodality of $\WHa$ seen in CALIFA data. The similarity with the bimodality observed in integrated galaxy spectra \citep{CidFernandes.etal.2011a} and the analogy of the concepts of hDIG and retired galaxy \citep{Stasinska.etal.2008a} strengthen the case for the proposed criteria, but it is the exact distribution of $\WHa$ in CALIFA spaxels that led to the quantitative criteria to separate the emission-line zones into hDIG/mDIG/SFc.  It is thus fit to start our analysis by checking whether this classification scheme also applies to MaNGA data.

%------------------------------- Figure -------------------------------%
\begin{figure*}
  \includegraphics[width=\textwidth, trim=36mm 8mm 36mm 10mm, clip]{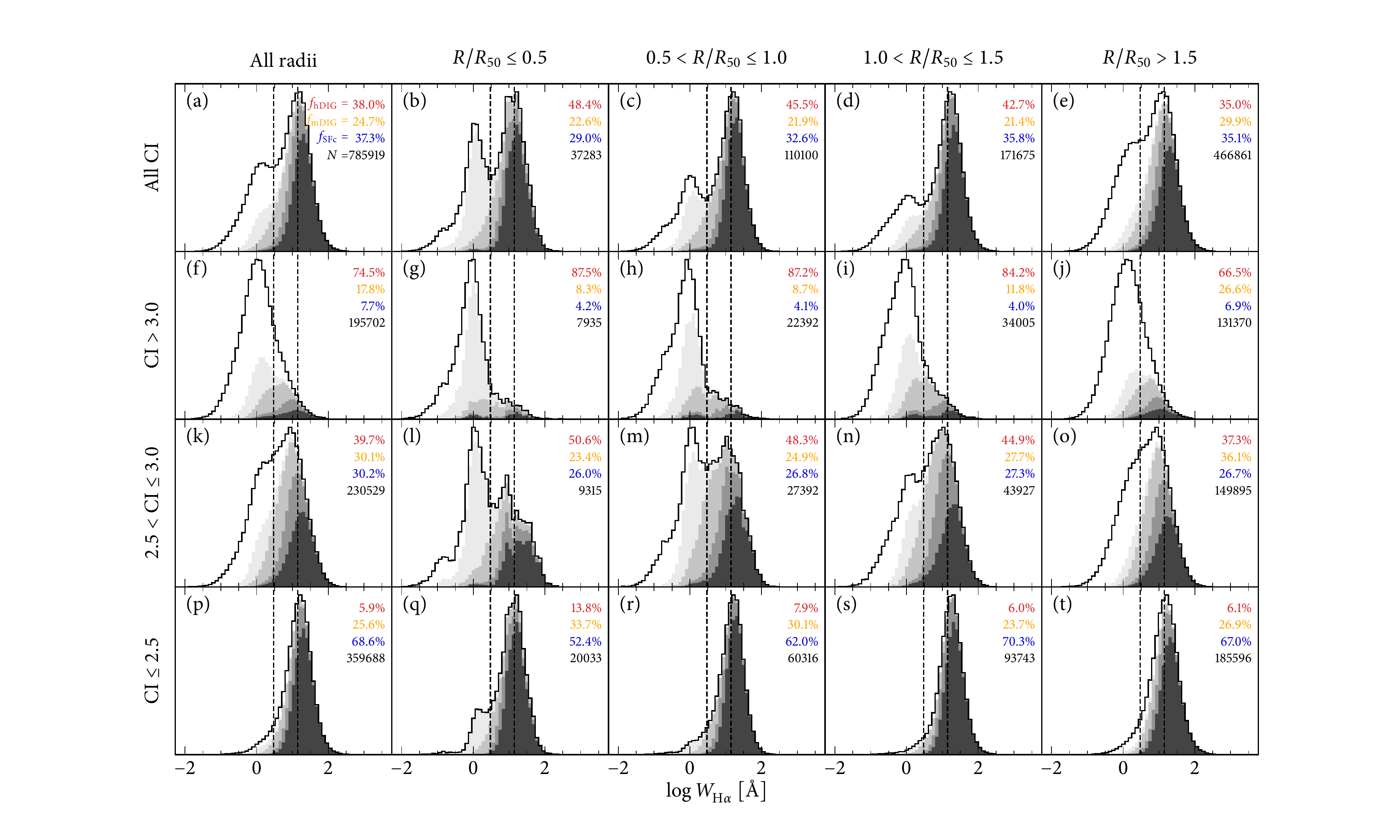}
  \caption{Distribution of $\WHa$ among 785\,919 spaxels of 3\,236
    MaNGA datacubes, broken into ranges of distance to the nucleus
    (2nd to 5th columns) and concentration index CI (2nd to 4th rows).
    The top left panel shows the full sample. Panels in the top row
    include all host morphologies, while the left column is for all
    radial distances. Dashed vertical lines at 3 and 14 \AA\ mark the
    boundaries of the HOLMES-powered DIG (hDIG, $\WHa < 3$ \AA), and
    star-forming complexes (SFc, $\WHa > 14$ \AA), as in
    \citetalias{Lacerda.etal.2018a}. Numbers on each panel show the
    average fraction of a galaxy's \Ha flux coming from spaxels in the
    $< 3$, $3$--$14$, and $> 14$ \AA\ ranges. The grey shading in the
    histograms represent cuts in \nii/\Ha: the lighter one is for the
    whole sample where \nii and \Ha are detected with $A/N \ge 2$,
    while the darker ones are for subsets of this $ A/N \ge 2$ sample
    where $\log\, \nii/\Ha < -0.10$, $-0.32$, and
    $-0.40$. }
\label{fig:WHa-hist}
\end{figure*}
%------------------------------- Figure -------------------------------%

Fig.~\ref{fig:WHa-hist} shows the distribution of $\WHa$ for MaNGA spaxels.  As in fig.~2 of \citetalias{Lacerda.etal.2018a}, the histograms are broken into ranges in distance from the nucleus ($R$) and CI, which can be used as a proxy for galaxy morphology \citep{Strateva.etal.2001a}.  Vertical dashed lines at $\WHa = 3$ and $14$ \AA\ mark the boundaries between hDIG, mDIG, and SFc. Numbers on each panel of Fig.~\ref{fig:WHa-hist} report the fraction of the total \Ha flux coming from the hDIG ($\WHa < 3$ \AA), mDIG (3--14 \AA), and SFc ($> 14$ \AA) components. We further subdivide each histogram into ranges in \nii/\Ha (grey histograms, which are commented later on). Here we concentrate on the full histograms (black solid line).

The top-left panel, which contains 785\,919 spaxels from our master sample of 3\,236 galaxies (from which we removed the 95\,429 spaxels with no \Ha detection), shows the evidently bimodal distribution of $\WHa$ previously seen both in CALIFA (\citetalias{Lacerda.etal.2018a}) and MaNGA \citep{Belfiore.etal.2016a}, as well as in the SDSS \citep{Bamford.etal.2008a, CidFernandes.etal.2011a}. Remarkably, the peak at $\WHa \sim 1$ \AA\ and the minimum at 3 \AA\ occur at the same places in these different datasets.  Stellar population analysis shows that line emission in this low-$\WHa$ population is predominantly due to photoionization by HOLMES \citep{Stasinska.etal.2008a, CidFernandes.etal.2011a}. The peak of the high-$\WHa$ population occurs at 14 \AA , close to $\WHa = 16$ \AA, the value found for integrated galaxy spectra \citep{CidFernandes.etal.2011a}, and the same value of $\WHa = 14$ \AA\ found for CALIFA \citepalias{Lacerda.etal.2018a}.
 
The behaviour of the $\WHa$ distribution with host morphology and distance from the nucleus is also very similar to that reported in \citetalias{Lacerda.etal.2018a}. As one progresses from early to late type morphologies the histograms gradually shift from hDIG to SFc dominated, as seen on the left column of  Fig.~\ref{fig:WHa-hist}. Early type galaxies (large CI, second row of plots) are hDIG-dominated throughout all radii. High $\WHa$ values among these galaxies occur mostly in their central regions, and are interpretable as due to AGN. Among disk galaxies the balance between the two $\WHa$ peaks reflects the bulge-to-disk ratio, with the ionizing radiation field changing from one dominated by old stars (HOLMES) in the bulge to one dominated by OB stars in the disk. The SFc population rises both as $R$ increases and as CI decreases.

To conclude, the similarity between the results found in this section and those reported in \citetalias{Lacerda.etal.2018a} justifies the use of the same $\WHa$-based hDIG/mDIG/SFc classification scheme to MaNGA data.

As an addition to the analysis, the histograms in Fig.~\ref{fig:WHa-hist} are shaded according to the \nii/\Ha ratio, darker shades corresponding to smaller \nii/\Ha.  The lightest shade corresponds to a selection of spaxels whose amplitude-to-noise $\geq 2$ in both \nii and \Ha. Darker shades draw histograms for the subsets of $A/N \geq 2$ spaxels where $\log\, \nii/\Ha < -0.10$, $< -0.32$, $< -0.40$, i.e.\ the optimal transposition of the \citet{Kewley.etal.2001a}, \citet{Kauffmann.etal.2003c} and \citet{Stasinska.etal.2006a} lines from the \Nii/\Ha versus \Oiii/\Hb plane to the WHAN (\Nii/\Ha versus \WHa) diagram \citep{CidFernandes.etal.2010a}. This shading scheme highlights the fact that regions with high \nii/\Ha tend to have low $\WHa$, suggesting that the higher values of \nii/\Ha  in the central zones is not a mere metallicity effect but due to ionization by HOLMES, which are more concentrated in the bulge of the galaxy than in the outskirts of the disc.
Without spatially resolved spectroscopy these DIG regions would thus bias the \nii/\Ha ratio with respect to what one would obtain by only considering regions dominated by star formation. The following sections are dedicated to quantifying this bias and investigating its effects upon estimates of the gas phase metallicity in star-forming galaxies.

%\natalia{\ojo Natalia: Remove this \\clearpage before submitting -- it is here to avoid compilation errors.}
%\clearpage

%**********************************************************************%
%                                                                      %
%                            Correction                                %
%                                                                      %
%**********************************************************************%
\section{How to correct integrated line luminosities for the DIG contribution}
\label{sec:howtocorrect}

We now use galaxies from our MaNGA master sample to estimate the contribution of the DIG to emission line fluxes.
We start by eliminating galaxies where the presence of AGN would significantly contribute to the line emission.  Photoionization models by \citet{Stasinska.etal.2006a}, which combined ionization by AGN and massive stars, have shown that the AGN contributes to $\lesssim 5$ per cent of the total \Ha emission for objects below the \citet{Kauffmann.etal.2003c} empirical demarcation line on the \Nii/\Ha versus \Oiii/\Hb plane. Therefore, selecting galaxies whose global integrated spectra are below this line, we define a SF MaNGA sample of 1\,409 galaxies.

Using the individual emission line measurements in each spaxel, we create several fictitious observations for the SF MaNGA galaxies by adding up individual fluxes within circular apertures of radius $R/R_{50}$. We call this total observed luminosity within the aperture $L_\mathrm{obs}$. We also calculate the observed equivalent width of \Ha within the aperture, $\WHa^\mathrm{obs}$, as the ratio of $L_\mathrm{obs}$ and the sum of the \Ha continua in those same spaxels.

For each of those $R < R_{50}$ circular apertures we add up the emission line fluxes of spaxels classified as SFc ($\WHa \ge 14$ \AA), thereby removing any contribution of hDIG and mDIG spaxels. The luminosity of an emission line coming only from SFc spaxels is dubbed $L_\mathrm{SFc}$.

In the following we use $L_\mathrm{SFc}$, $L_\mathrm{obs}$ and $\WHa^\mathrm{obs}$ to guide our correction of emission line luminosities by the DIG contribution in single-aperture observations.

%**********************************************************************%
\subsection{Correcting entire galaxies}
\label{sec:entire}

%------------------------------- Figure -------------------------------%
\begin{figure*}
  %trim=g b d h,
  \includegraphics[width=.85\textwidth, trim=0 .5cm 0 0]{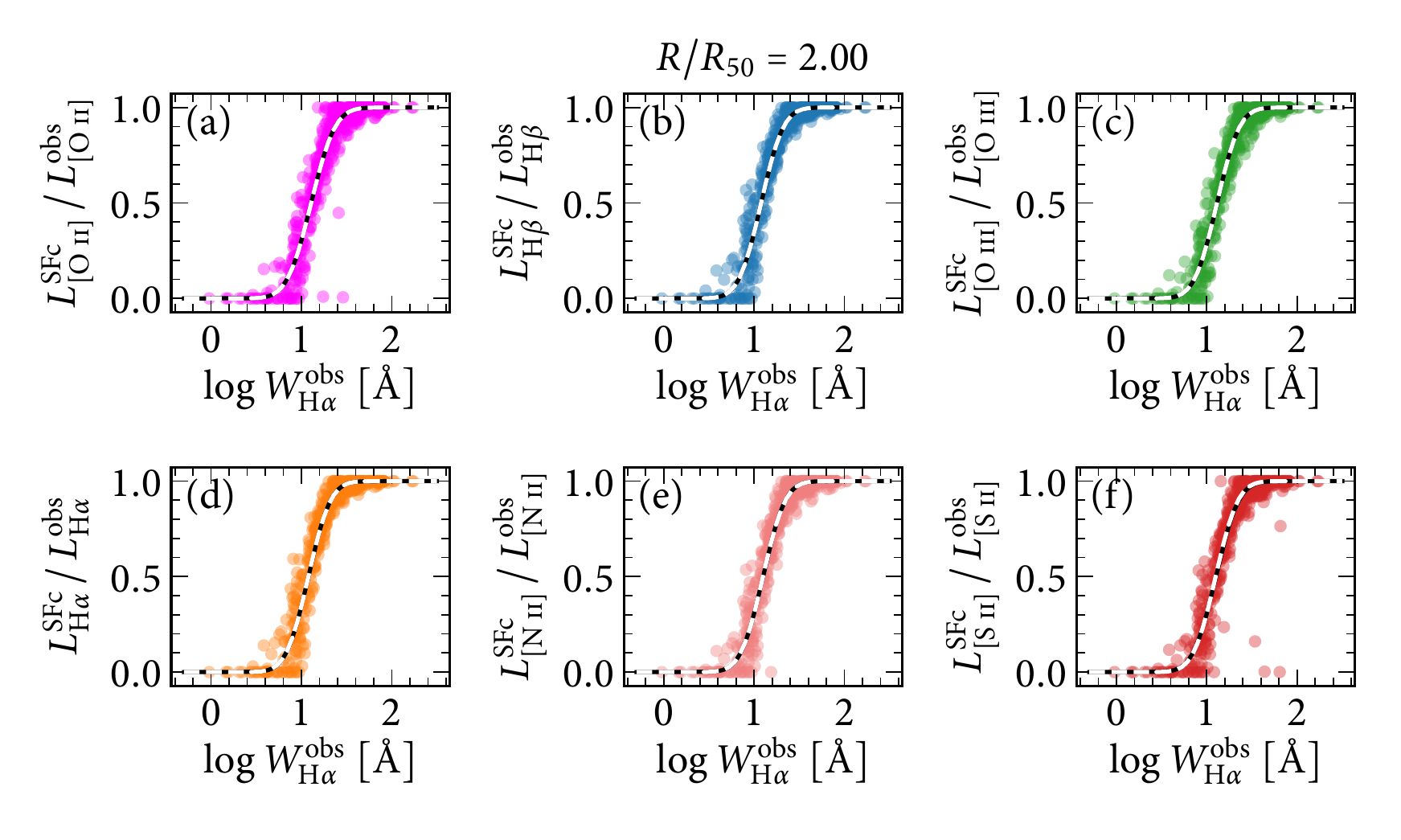}
  \caption{The ratio of the line luminosity in SFc spaxels
    $L_\mathrm{SFc}$ to the total luminosity $L_\mathrm{obs}$ as a
    function of $\WHa^\mathrm{obs}$. All quantities are
    measured in a circular aperture of size $R/R_{50} = 2.0$ for 504  MaNGA SF galaxies. Each
    panel shows the results for a given emission line. The overimposed 
    solid black and dashed white lines show a fit to the data using the
    function defined by equation~\ref{eq:erf}. This fit defines the
    correction to emission lines appropriate for entire galaxies, and
    fitted parameters are given in Table~\ref{tab:cor-entire}. 
  }
\label{fig:fcor}
\end{figure*}
%------------------------------- Figure -------------------------------%

In order to find a correction procedure as simple as possible, we searched for the best observational parameter with which to link the value of $f \equiv L_\mathrm{SFc}/L_\mathrm{obs}$. We experimented several easily accessible parameters measured within a circular aperture: stellar mass, the 4000-\AA\ break and $\WHa^\mathrm{obs}$ (where the superscript `obs' reinforces the fact that this quantity was measured within an aperture, and is not the \WHa measured in a spaxel as in Fig.~\ref{fig:WHa-hist}). It turned out that $\WHa^\mathrm{obs}$ was the one which gave the smallest dispersion. This is not really surprising, as our definition of the DIG is closely related to \WHa.

Fig.~\ref{fig:fcor} shows $L_\mathrm{SFc}/L_\mathrm{obs}$ as a function of $\WHa^\mathrm{obs}$ for \oii, \Hb, \oiii, \Ha, \nii and \sii\footnote{We use the following notations: \oii = $\oii\lambda 3726
  + \oii\lambda 3729$, \sii = $\sii\lambda 6716 + \sii\lambda 6731$, \oiii = \Oiii and \nii = \Nii.}.  The plot shows all the 504 galaxies for which we could calculate circular apertures of size $R/R_{50} = 2.0$, i.e.\ which had all spaxels within $2 R_{50}$ unmasked. Note that the figure is quasi identical for any value of $R/R_{50}$ larger than $\sim 1.0$.

On each panel, the dashed line shows the fit to the data using a normal cumulative distribution function on $x = \log \WHa$:
\begin{equation}
  \label{eq:erf}
  f(\log \WHa^\mathrm{obs}, x_0, \sigma) =
    \frac{1}{\sqrt{2\pi} \sigma} \int_{-\infty}^{\log \WHa^\mathrm{obs}}
    e^{ - (x-x_0)^2 / 2 \sigma^2} dx.
\end{equation}
The fitted parameters are $x_0$, the centre of the sloping curve, and $\sigma$, the width of the underlying Gaussian (thus a larger $\sigma$ means a shallower slope). The fitted parameters for each emission line are given in Table~\ref{tab:cor-entire}.

Using those parameters, one can correct the emission line luminosities in single-aperture spectra of \emph{entire galaxies}.  The corrected SF emission is given by \begin{equation}
  L_\mathrm{SFc} = L_\mathrm{obs} \frac{L_\mathrm{SFc}}{L_\mathrm{obs}}
                = L_\mathrm{obs}\, f(\log \WHa^\mathrm{obs}, x_0, \sigma).
\end{equation}
So to correct an observed line luminosity $L_\mathrm{obs}$, one simply needs to choose the appropriate values of $x_0, \sigma$ for that given emission line and to measure $\log \WHa^\mathrm{obs}$.  In Appendix~\ref{sec:code}, we give a sample code to correct emission line luminosities. We warn that this correction should not be applied to galaxies with $\WHa \lesssim 10$~\AA, since they are dominated by hDIG and mDIG emission (note how most curves of $L_\mathrm{SFc}/L_\mathrm{obs}$ in Fig.~\ref{fig:fcor} are close to zero when $\WHa^\mathrm{obs} =10$~\AA).

A caveat to bear in mind is that, although MaNGA currently offers the largest homogeneous IFS data for this kind of study, its $\sim$kpc linear resolution limits us to identifying spaxels dominated by star forming complexes using the rather arbitrary cut at $\WHa = 14$ \AA. If a statistically relevant sample of SF galaxies with higher resolution becomes available, those decisions and cuts should be revisited.

\begin{table}
    \centering
    \caption{Correction parameters suited for entire galaxies.}
    \begin{tabular}{lccc}
      \hline
      Line                   & $x_0$ [log \AA] & $\sigma$ [log \AA] \\
      \hline
%if cor == 2.0:
      $\oii\lambda3726+3729$ & $1.108$ & $0.2086$ \\
      \Hb                    & $1.078$ & $0.1944$ \\
      \Oiii                  & $1.113$ & $0.1963$ \\
      \Ha                    & $1.077$ & $0.1932$ \\
      \Nii                   & $1.098$ & $0.1944$ \\
      $\sii\lambda6716+6731$ & $1.100$ & $0.1993$ \\
      \hline
    \end{tabular}
    \label{tab:cor-entire}
\end{table}

\begin{table}
    \centering
    \caption{Correction parameters suited for single-aperture observations of central parts of galaxies.}
    \begin{tabular}{lccc}
      \hline
      Line                   & $x_0$ [log \AA] & $\sigma$ [log \AA] \\
      \hline
%if cor == 0.7:
      $\oii\lambda3726+3729$ & $1.130$ & $0.1302$ \\
      \Hb                    & $1.121$ & $0.1341$ \\
      \Oiii                  & $1.131$ & $0.1262$ \\
      \Ha                    & $1.121$ & $0.1340$ \\
      \Nii                   & $1.131$ & $0.1300$ \\
      $\sii\lambda6716+6731$ & $1.130$ & $0.1316$ \\
      \hline
    \end{tabular}
    \label{tab:cor-aperture}
\end{table}
%**********************************************************************%

%**********************************************************************%
\subsection{Accounting for aperture effects}
\label{sec:aperture}

%------------------------------- Figure -------------------------------%
\begin{figure*}
  \includegraphics[width=.85\textwidth, trim=0 .5cm 0 0]{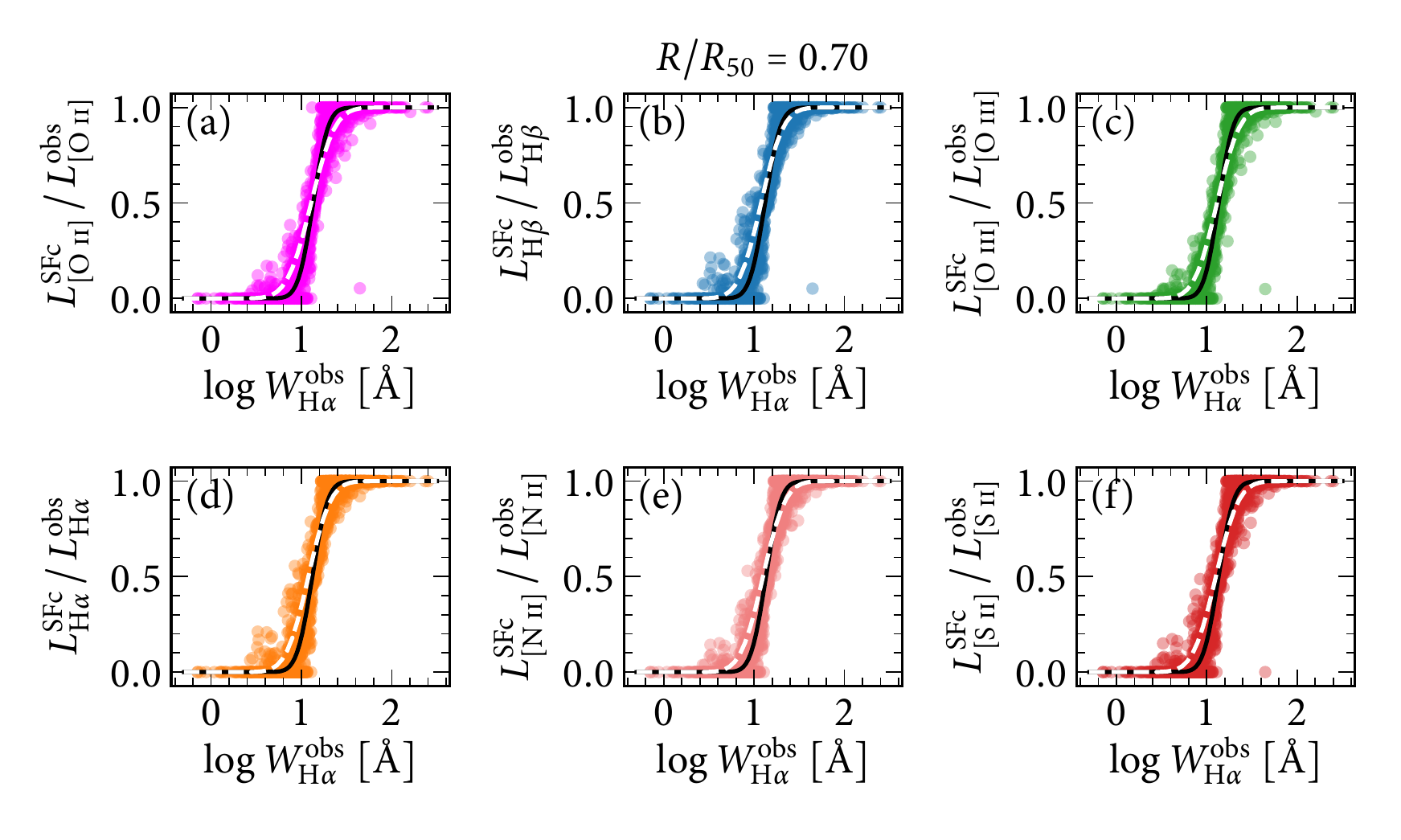}
  \caption{As Fig.~\ref{fig:fcor}, but for a circular
      aperture of size $R/R_{50} = 0.7$ for 1\,409  MaNGA SF galaxies. The solid black  line is a fit to
      those measurements, and the dashed white  line repeats the fit
      for $R/R_{50} = 2.0$ from Fig.~\ref{fig:fcor} for comparison.
      The corrections shown in those plots are appropriate for central
      observations of galaxies, such as those in the SDSS DR7, and fitted parameters are given in
      Table~\ref{tab:cor-aperture}.
    }
\label{fig:fcorap}
\end{figure*}
%------------------------------- Figure -------------------------------%

As shown in \citetalias{Lacerda.etal.2018a} and also in Fig. \ref{fig:WHa-hist},  the importance of the DIG varies with the galactocentric distance. It is thus expected that, in single aperture observations, the corrections for the DIG should depend on the covering fraction of the aperture. We therefore repeated the procedure from Section \ref{sec:entire} for various values of  $R/R_{50}$.

It turns out that the fitted curves for $f(\log \WHa^\mathrm{obs}, x_0, \sigma)$ change very little from $R/R_{50} = 0.4$ to $0.8$. We thus show in Fig.~\ref{fig:fcorap} only the case corresponding to $R/R_{50} = 0.7$. Table~\ref{tab:cor-aperture} gives the corresponding values of $x_0$ and $\sigma$ for each emission line.

In the next two sections, we apply these corrections to a sample of SDSS galaxies suitable for emission-line abundance analyses. 

%**********************************************************************%

%**********************************************************************%
%                                                                      %
%                       Metallicity estimates                          %
%                                                                      %
%**********************************************************************%
\section{Effect of the DIG on metallicity estimates in SDSS galaxies
}
\label{sec:metallicity estimates}

%**********************************************************************%
%                                                                      %
%                          SDSS data                                   %
%                                                                      %
%**********************************************************************%
\subsection{The SDSS DR7 sample }
\label{sec:SDSS}

We select galaxies observed by the SDSS DR7 \citep{Abazajian.etal.2009a} from the \starlight database\footnote{The public database can be accessed at \protect\url{http://starlight.ufsc.br}.}  \citep{CidFernandes.etal.2005a, CidFernandes.etal.2009a}.  We apply a flux renormalisation correction to match the fibre and spectral $r$-band photometries using the \texttt{spectofibre} factor provided by the MPA-JHU team\footnote{\url{http://www.mpa-garching.mpg.de/SDSS/DR7/raw_data.html}}.  The stellar populations models used are from \citet{Bruzual.Charlot.2003a} as described in \cite{Asari.etal.2007a}. The stellar mass $M$ is calculated from the \starlight spectral synthesis modelling as in \citet{CidFernandes.etal.2005a}, corrected for the mass outside the fibre using the $z$-band photometry.  Line luminosities are measured in the residual spectra after removing the stellar continua from the observed spectra using the method detailed in \citet{Mateus.etal.2006a, Stasinska.etal.2006a}.

We start from a sample of 777\,967 galaxies similar to the one described by \cite{KozielWierzbowska.etal.2017a}: we select all galaxies  from the Main Galaxy Sample \citep{Strauss.etal.2002a} and the Luminous Red Galaxy \citep{Eisenstein.etal.2001a} catalogues; we impose a minimum signal-to-noise of 5 at the 4020 \AA\ continuum, and we remove galaxies with $M \le 10^7$~M$_\odot$ or with negative values of $R_{50}$.

From this starting sample, we obtain a sample of 94\,335 SF galaxies by applying the following criteria. Inspired by \cite{Mannucci.etal.2010a}, we select objects (1) below the \citet{Kauffmann.etal.2003c} line on the \Nii/\Ha versus \Oiii/\Hb plane, (2) which have a minimum $S/N$ of 25 in their \Ha line detection, (3) with nebular $A_V < 2.5$ (calculated from the Balmer decrement using a \citet{Cardelli.Clayton.Mathis.1989a} extinction law with $R_V = 3.1$), and (4) with redshifts limited to the $0.07 \le z \le 0.30$ range. We also impose other common sense cuts: we remove very edge-on systems by selecting $b/a \ge 0.3$, and remove hDIG/mDIG-dominated spectra by leaving only those with $\WHa \ge 10$ \AA.

\subsection{Metallicity biases }
  \label{sec:metallicitybias}
  
%------------------------------- Figure -------------------------------%
\begin{figure*} 
  \includegraphics[width=.75\textwidth, trim=0 2.5cm 2.5cm 0]{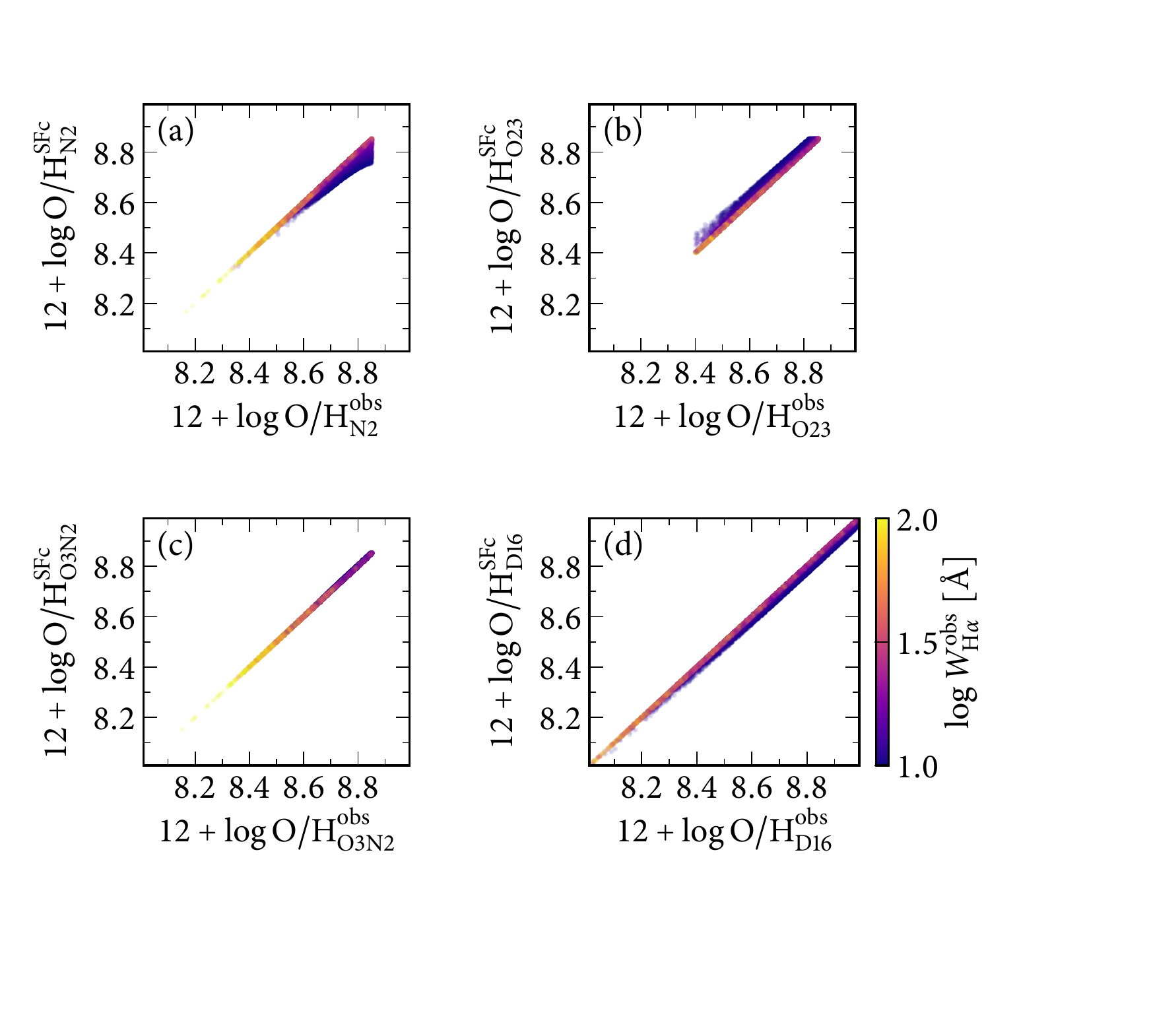}
  \caption{Comparison of nebular metallicity estimates for 94\,335 SF galaxies from the SDSS DR7. The abscissa shows metallicities calculated using emission lines with no correction, whereas the ordinate is after removing the DIG emission using the correction appropriate for SDSS-like apertures. Each panel shows a different calibration applied with different indices: N2, O23 and O3N2 (using calibrations from \citealp{Curti.etal.2017a}) and D16 (by \citealp{Dopita.etal.2016a}, whose index is a combination of N2 and N2S2). Points are colour-coded by their $\WHa$, to highlight where the DIG correction is more important.} 
\label{fig:Z}
\end{figure*}
%------------------------------- Figure -------------------------------%
  
We use the recent \citet{Curti.etal.2017a} calibrations to estimate the gas-phase oxygen abundances (O/H) from three strong line indices: N2, O23 and O3N2\footnote{We use the notation N2 for \Nii/\Ha, O23 for (\oiii$\lambda$5007+4959 + \oii$\lambda$3726+3729)/\Hb\ -- which is called R23 by \citet{Curti.etal.2017a} --, and O3N2 for (\Oiii/\Hb)/(\Nii/\Ha).}, which are valid in the $12 + \log \mathrm{O/H} =$ 7.60--8.85, 8.40--8.85, and 7.60--8.85 ranges, respectively.  We also use the \citet{Dopita.etal.2016a} calibration of O/H as a function of N2 and N2S2\footnote{N2S2 is for \Nii/\sii$\lambda$6716+6731.} (their equations 1 and 2, valid in the $12 + \log \mathrm{O/H} =$ 8.00--9.00 range), and call it D16.  We correct all the relevant emission lines to calculate each index, and for O23 we apply a reddening correction to all lines using the \Ha/\Hb Balmer decrement and a \citet{Cardelli.Clayton.Mathis.1989a} dust law with $R_V = 3.1$.

Fig.~\ref{fig:Z} compares O/H obtained from the observed emission lines to O/H calculated from the emission lines after applying the correction for DIG emission assuming a covering fraction $R/R_{50} = 0.7$, thus applicable to SDSS observations of galaxies up to redshifts of $\sim 0.5$ (which is roughly the redshift limit for objects in the Luminous Red Galaxy sample of SDSS DR7).

It can be seen that the strongest biases are found at the highest metallicities.  O3N2 is the least affected abundance indicator, since both \Oiii/\Hb and \Nii/\Ha rise almost in tandem in the DIG (as obvious from the position of DIG and SFc spaxels in the \oiii/\Hb versus \nii/\Ha diagram, e.g.\ see fig.~9 of \citetalias{Lacerda.etal.2018a}).  \citet{Kumari.etal.2019a} also concluded that O3N2 is the abundance indicator that is the least biased by the DIG.  The N2 index is the most affected by the DIG, especially at high metallicity, leading to an overestimate of O/H by up to 0.1 dex if no correction is made.
 
The points in Fig.~\ref{fig:Z} have been colour-coded according to their value of $\WHa^\mathrm{obs}$, as indicated in the colour scale on the bottom right. As expected, it is those galaxies with the smallest values of $\WHa^\mathrm{obs}$ which show the largest abundance bias.  
The fact that galaxies with high metallicities are also the ones with low $\WHa^\mathrm{obs}$ is related to those objects having both a larger DIG emission and a smaller specific SFR (due to the mass--metallicity relation and the decrease in specific SFR with increasing stellar mass, e.g. \citealp{Kauffmann.etal.2003c}).
Moreover, the low $\WHa^\mathrm{obs}$--high O/H (thus high stellar mass) relation is also in part due to observational selection effects: at a given value of $\WHa^\mathrm{obs}$, galaxies with the highest stellar mass are brighter and more easily detectable. Since we are simply showing the effect of the DIG to O/H estimates, these selection effects are not key to this paper.
Note that, for higher signal-to-noise observations, which will be obtained for surveys using larger telescopes than the one used for SDSS, a larger proportion of galaxies with low values of $\WHa^\mathrm{obs}$ will be present, because the condition of a high signal-to-noise in the emission lines will be more easily reached. Thus, the average abundance bias will be larger than for the SDSS survey.

%**********************************************************************%
%                                                                      %
%                               MZSFR                                  %
%                                                                      %
%**********************************************************************%
\section{Effect of the DIG on the $M$--$Z$--SFR relation}
\label{sec:MZSFR}

The dependence of the $M$--$Z$ relation on the SFR is not yet clear.  Previous studies using integrated spectra showed that nearby galaxies in a fixed stellar mass bin show an anti-correlation between $Z$ and SFR at low values of $M$ \citep{Ellison.etal.2008a, LaraLopez.etal.2010a, LaraLopez.etal.2010b, Mannucci.etal.2010a}, suggesting e.g.\ inflow of low-$Z$ gas to feed SF in the galaxies, or outflows of enriched gas.  \citet{SanchezAlmeida.SanchezMenguiano.2019a}, using a sample of 736 nearby spiral galaxies with MaNGA spatially resolved data, suggested that the global $M$--$Z$--SFR relation is a result of the local anti-correlation between the SFR surface density and the gas metallicity. Nevertheless, other studies, using spatially resolved data and various metallicity indicators, could not find such relation between the $M$--$Z$ and SFR, indicating that the relation may arise from e.g.\ an aperture effect in single-fibre spectroscopy \citep{Sanchez.etal.2017a} or how the $M$-bins are defined \citep{BarreraBallesteros.etal.2017a}. The $M$--$Z$--SFR relation also depends on the choice of the metallicity indicator, as demonstrated by \citet{Kashino.etal.2016b} and as can be seen in Fig.~\ref{fig:MZSFR} below. Making use of the calibration proposed by \citet{Dopita.etal.2016a}, those authors showed that the anti-correlation between metallicity and SFR disappears for the lowest masses.
 
All things considered, correctly estimating the gas-phase oxygen abundances of galaxies is crucial in the analysis of such relations. As mentioned above, the DIG contribution to the line emission is not negligible and will affect the metallicity indicators if not taken into account.

%------------------------------- Figure -------------------------------%
\begin{figure*}
  \centering
  \includegraphics[width=.93\textwidth, trim=2cm 1.5cm 8cm 2cm, clip]{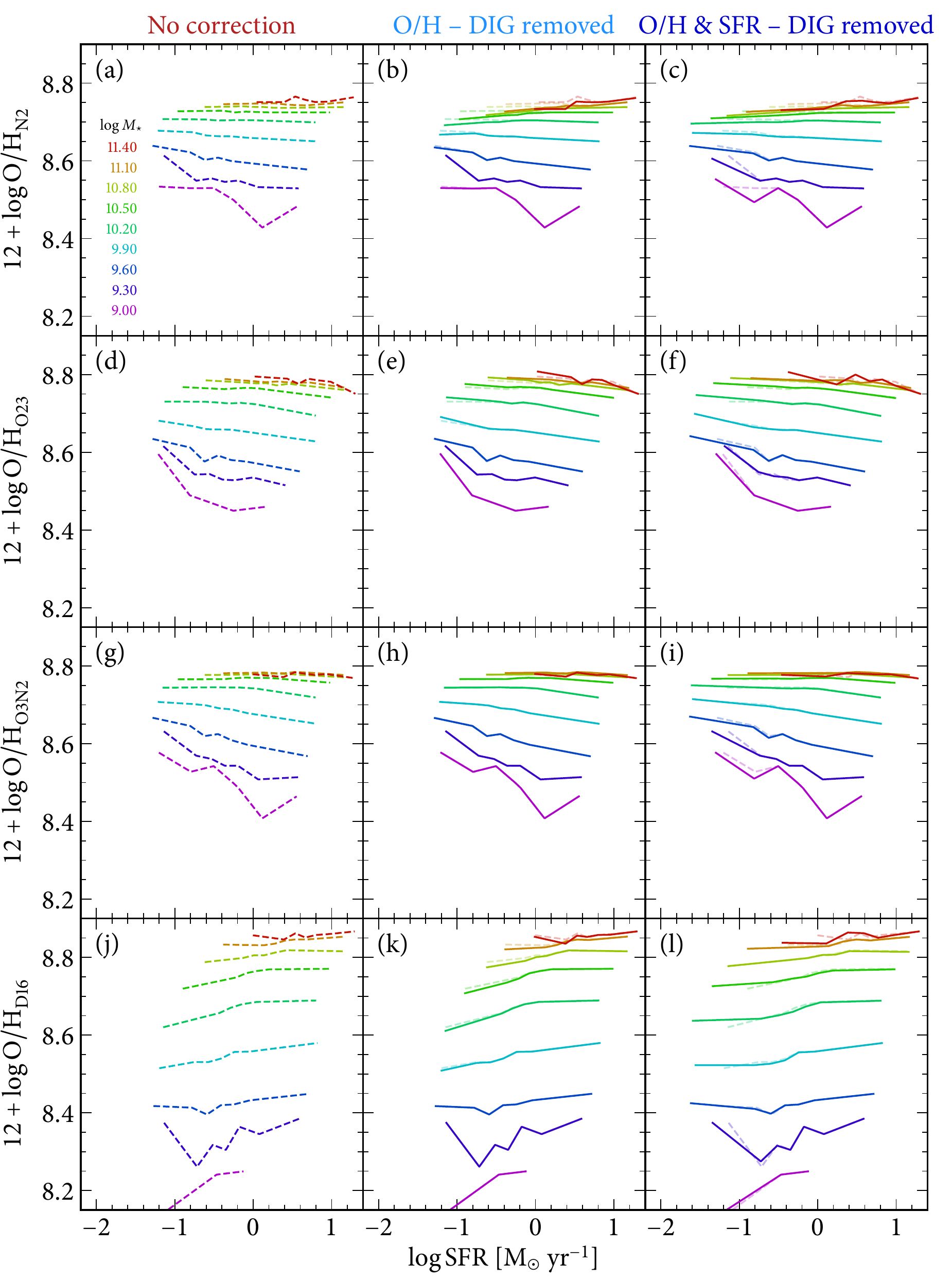}
  \caption{Each panel shows O/H as a function of SFR in
    $0.30$-dex-wide stellar mass bins for a sample of 94\,335 SF galaxies from the SDSS DR7. Panel (a) shows the $\log M$ centre values for each bin, and also the colour-coding for each
    median curve.  Rows show results for the four different O/H
    calibrations discussed in Sec.~\ref{sec:metallicitybias}.  The
    column of panels on the right shows the observed relation (dashed lines); the
    middle corrects the emission lines for the O/H computation by the
    DIG emission (solid lines); and the right column also corrects the SFR for the
    DIG contribution (solid lines). The `no correction' relation shown on the left
    column as dashed lines is repeated on the two other columns for comparison. }
\label{fig:MZSFR}
\end{figure*}
%------------------------------- Figure -------------------------------%

Fig.~\ref{fig:MZSFR} shows the SFR versus O/H in bins of stellar mass, as in fig.~1 of \citet{Mannucci.etal.2010a}. Each row shows a different estimate of O/H using the four indices from the last section. On the left we show the plots where O/H and SFR have been obtained without any DIG correction. In the middle, we show the same quantities, but now correcting the determination of O/H for the presence of DIG. On the right we again show the same quantities, but now correcting both O/H and SFR for the presence of DIG. The dashed lines from the left panels are repeated in the middle and right panels, which allows one to compare more accurately the `no correction' with the `DIG removed' results.

The SFR is calculated from the \Ha luminosity from equation~(8) of \citet{Asari.etal.2007a}, which was calibrated for solar metallicity and a \citet{Chabrier.2003a} IMF between 0.1 and 100 M$_\odot$ using \citet{Bruzual.Charlot.2003a} stellar population models:
\begin{equation}
  \label{eq:SFR_LHa}
  \mathrm{SFR} 
  = 2 \,\mathrm{M}_\odot\, \mathrm{yr}^{-1} \left(
    \frac{L_{\Ha}}{10^8 \,\mathrm{L}_\odot} \right).
\end{equation}
The underlying assumptions under this conversion are that (1) the SFR is constant over the lifetime of massive ionizing stars ($\sim 10$~Myr), (2) the nebulae are ionization-bounded, i.e.\ no ionizing photons escape, (3) the case B recombination applies, (4) all \Ha luminosity is produced by the ionization of massive stars, i.e.\ it ignores other ionization sources such as HOLMES.

The effect of the DIG on the SFR estimate is to increase the SFR with respect to its real value because some of the \Ha emission is attributed to star formation while it is actually due to other ionization sources (such as HOLMES). This effect is stronger for the most massive galaxies, where a significant fraction of the \Ha emission is not due to star formation. 

As regards the effect of the DIG on O/H in Fig.~\ref{fig:MZSFR}, it seems negligible when using the O3N2 or the O23 indices (although in the latter case this index is almost independent of metallicity for $12 + \log \mathrm{O/H}$ between 8.0 and 8.4, see e.g.\ \citealp{Curti.etal.2017a}).  The effect however becomes noticeable when using the D16 index, and even more so when using N2. In the latter case, O/H becomes correlated with SFR for high mass galaxies instead of being flat, because the correction is greater at smaller values of \WHa (which corresponds to smaller SFR at a given mass).

We must note that the DIG-corrected abundances depend on the indices used to derive them -- especially at the highest metallicities and lowest SFRs. This indicates that the strong line calibrations we used are slightly biased, probably precisely because they did not consider the influence of the DIG. This issue needs to be further investigated. 

Note that the effects found in this paper could actually be much greater, and could also affect low-mass galaxies, as one may notice on the right column of Fig.~\ref{fig:MZSFR}, mainly regarding N2 and O3N2 indices. To check this we would need higher resolution data, say MUSE-like observations for the same set of MaNGA galaxies. One should beware, however, that data with too high resolution should also be used with care, since the reasoning behind the \WHa hDIG/mDIG/SFc classification is founded on the fact that the ionizing source and the ionized gas are observed on the same spaxel.

The correction for DIG using \WHa for MaNGA galaxies is however physically motivated. As argued by \citetalias{Lacerda.etal.2018a}, \WHa is linked to the ionizing stellar population, ranging from $1$~\AA\ for HOLMES to several hundred \AA\ for ionizing young populations. \WHa is also an intensive quantity, i.e.\ observing two hDIG regions into the same line of sight yields a resulting \WHa which is characteristic of hDIG regions. This is in contrast to a cut in \Ha surface brightness, which is an extensive quantity: two DIG-like regions overimposed in the same line of sight will have a higher $\Sigma_{\Ha}$ and may cross into the SFc territory. This can be especially important for bulges in the central parts of galaxies, exactly the parts observed by SDSS. As a matter of curiosity, we have tested how a cut in $\Sigma_{\Ha}$ would affect the $M$--$Z$--SFR relation. A simple cut like the one proposed by \citet{Zhang.etal.2017c} results in similar shifts in the $M$--$Z$--SFR for all calibrations, but affecting all mass bins. The only major difference to our results is an overall upward shift in D16 metallicities. We have also tested hybrid methods, which employ DIG removal criteria based on $\Sigma_{\Ha}$ and \sii/\Ha, such the ones proposed by \citet{Kaplan.etal.2016a} and \citet{Poetrodjojo.etal.2019a}. The results are very similar to the ones with a simple $\Sigma_{\Ha}$, but much noisier. Since the $\Sigma_{\Ha}$ and/or \sii/\Ha cuts change from galaxy to galaxy, we did not find a global parameter which would be tightly correlated to the emission line correction factors. Cuts in \sii/\Ha also run against the logic of this work, which is quantifying the impact of the DIG on emission line fluxes, and propagating it to line ratios used as indices for metallicity calibrations: using a limit in \sii/\Ha could be equivalent to forcing a solution from the outset for indices based on \sii/\Ha itself.

There is an added complication that we have not dealt with: the original $M$--$Z$--SFR relations \citep{Ellison.etal.2008a, LaraLopez.etal.2010a, LaraLopez.etal.2010b, Mannucci.etal.2010a} were found for O/H and SFR computed using line luminosities measured inside the fibre. For O/H, since metallicity gradients tend to be very small, an aperture correction may be minimal. For the SFR, one might be able to correct to the total SFR using a method similar to \citet{Brinchmann.etal.2004a}. \citet*{Yates.Kauffmann.Guo.2012a}, for instance, have found that the median total SFR is about 0.6 dex higher than the fibre SFR, with the effect being larger for low-mass (thus less concentrated), low-redshift galaxies. Correcting the SFR in a satisfactory manner for aperture effects is however out of the scope of this work, since any estimate (either trying to account for \Ha outside the fibre or correcting the integrated spectrum for the light outside the fibre and modelling the stellar population synthesis to obtain the SFR, as outlined in section 5 of \citealp{Asari.etal.2007a}) would be severely affected by dust variations from one region to the other within a galaxy (Vale Asari \etal, in preparation).

%**********************************************************************%
%                                                                      %
%                               Summary                                %
%                                                                      %
%**********************************************************************%
\section{Summary}
\label{sec:summary}

The presence of the DIG has long been suspected to affect integrated spectra of galaxies.  In this paper, 
we have focused on the effect of the DIG on abundance determinations in emission-line galaxies.

Defining the DIG according to \WHa, as shown to be appropriate by \citetalias{Lacerda.etal.2018a}, we have used a set of 1\,409 SF galaxies from the IFS MaNGA survey to find what percentage of the various emission lines luminosities is due to the DIG. Our method relies on high-S/N data from SFc spaxels, instead of relying on measurements of emission line in noisier, DIG-dominated spaxels.

We have thus proposed a method to recover the line luminosities arising from SFc regions alone for galaxies whose observed \WHa are larger than about 10\AA\ (galaxies or galaxy zones with \WHa smaller than such a value are dominated by mDIG/hDIG emission and cannot be treated as star-forming galaxies). Since the importance of the DIG depends on the galactocentric distance, we propose two sets of corrections: one for integrated spectra of entire galaxies, whose number is bound to increase drastically with the development of deep and ultra-deep surveys; and another for single aperture observations of galaxies in the local universe, such as the SDSS DR7. 

We apply this DIG correction to a sample of 94\,335 SF galaxies from the SDSS DR7. Among the indices usually used to estimate the oxygen abundance, it is N2 which is the most affected by the presence of the DIG, producing a bias of 0.1 dex in O/H at the highest metallicity and lowest \WHa. On the other hand, O3N2 is not much affected (in agreement with \citealp{Kumari.etal.2019a}), whereas O23 and D16 are somewhat biased.

We have then revisited the $M$--$Z$--SFR relation after correcting for the DIG on both O/H and SFR. The effect is noticeable for the highest stellar masses and metallicities, and may result in O/H \emph{increasing} with SFR at high stellar mass (especially when using N2). This effect of course is based on a correction made with IFS data with 1--2 kpc resolution, so it should be taken as a cautionary stepping stone. The $M$--$Z$--SFR relation might have changed more dramatically if we had at hand a large dataset of higher-resolution IFS data, and the effect might have been noticeable even for low-mass galaxies.

Note that the so-called fundamental $M$--$Z$--SFR relation of \citet{Mannucci.etal.2010a}, which was obtained ignoring the DIG and taking the average of two metallicity indices, was also based on SDSS galaxies with S/N in \Ha larger than 25. This criterion removes preferentially massive galaxies ($\log M \gtrsim 11$ M$_\odot$). A $M$--$Z$--SFR relation based on selection criteria less affected by observing conditions may look quite different. Also, $Z$ and SFR were obtained with line measurements inside a spectral fibre. An aperture correction can be very tricky if one wants to take into account line luminosity variations within a galaxy (Vale Asari \etal, in preparation) and outside the fibre, which is especially important to correct the SFR. Attempts to tackle this issue have been made by other authors \citep[e.g.][]{Yates.Kauffmann.Guo.2012a}.

Given how important the $M$--$Z$--SFR relation is for galaxy chemical evolution models, all those selection, observational, and analysis caveats should be taken into account when comparing this empirical relation to models \citep[e.g.][]{Yates.Kauffmann.Guo.2012a,Kashino.etal.2016b}.  The changes brought about by the removal of the DIG contamination reveal that all assumptions underlying this relation must be carefully considered.

%**********************************************************************%
%                                                                      %
%                           Acknowledgements                           %
%                                                                      %
%**********************************************************************%
\section*{Acknowledgements}
    We are thankful to the referee, Brent Groves, for the useful comments.
    NVA acknowledges support of the Royal Society and the Newton Fund via the award of a Royal Society--Newton Advanced Fellowship (grant NAF\textbackslash{}R1\textbackslash{}180403), and of Funda\c{c}\~ao de Amparo \`a Pesquisa e Inova\c{c}\~ao de Santa Catarina (FAPESC) and Conselho Nacional de Desenvolvimento Cient\'{i}fico e Tecnol\'{o}gico (CNPq). 
    This study was financed in part by the Coordena\c{c}\~ao de Aperfei\c{c}oamento de Pessoal de N\'{\i}vel Superior -- Brasil (CAPES) -- Finance Code 001.
    NVA, RCF, GS, ALA, AW and TZF acknowledge the support from the CAPES CSF--PVE project 88881.068116/2014-01. 
    GSC  acknowledges the support  by  the  Comit\'{e}  Mixto ESO-Chile and the DGI at University of Antofagasta.
    AW acknowledges financial support from Funda\c{c}\~ao de Amparo \`a Pesquisa do Estado de S\~ao Paulo (FAPESP) process number 2019/01768-6.
    ALA and DRD acknowledge the financial support from Programa Nacional de P\'{o}s Doutorado (PNPD/CAPES).
    RCF acknowledges financial support from CNPq.
    The authors thank Vivienne Wild, Gustavo Bruzual and St\'ephane Charlot for their help with the correct references for the updated spectral synthesis models.

    Funding for the SDSS and SDSS-II has been provided by the Alfred P. Sloan Foundation, the Participating Institutions, the National Science Foundation, the U.S. Department of Energy, the National Aeronautics and Space Administration, the Japanese Monbukagakusho, the Max Planck Society, and the Higher Education Funding Council for England. The SDSS Web Site is http://www.sdss.org/.
    The SDSS is managed by the Astrophysical Research Consortium for the Participating Institutions. The Participating Institutions are the American Museum of Natural History, Astrophysical Institute Potsdam, University of Basel, University of Cambridge, Case Western Reserve University, University of Chicago, Drexel University, Fermilab, the Institute for Advanced Study, the Japan Participation Group, Johns Hopkins University, the Joint Institute for Nuclear Astrophysics, the Kavli Institute for Particle Astrophysics and Cosmology, the Korean Scientist Group, the Chinese Academy of Sciences (LAMOST), Los Alamos National Laboratory, the Max-Planck-Institute for Astronomy (MPIA), the Max-Planck-Institute for Astrophysics (MPA), New Mexico State University, Ohio State University, University of Pittsburgh, University of Portsmouth, Princeton University, the United States Naval Observatory, and the University of Washington.

    Funding for the Sloan Digital Sky Survey IV has been provided by the Alfred P. Sloan Foundation, the U.S. Department of Energy Office of Science, and the Participating Institutions. SDSS-IV acknowledges support and resources from the Center for High-Performance Computing at the University of Utah. %The SDSS web site is www.sdss.org.
    SDSS-IV is managed by the Astrophysical Research Consortium for the 
Participating Institutions of the SDSS Collaboration including the 
Brazilian Participation Group, the Carnegie Institution for Science, 
Carnegie Mellon University, the Chilean Participation Group, the French Participation Group, Harvard-Smithsonian Center for Astrophysics, 
Instituto de Astrof\'isica de Canarias, The Johns Hopkins University, Kavli Institute for the Physics and Mathematics of the Universe (IPMU) / 
University of Tokyo, the Korean Participation Group, Lawrence Berkeley National Laboratory, 
Leibniz Institut f\"ur Astrophysik Potsdam (AIP),  
Max-Planck-Institut f\"ur Astronomie (MPIA Heidelberg), 
Max-Planck-Institut f\"ur Astrophysik (MPA Garching), 
Max-Planck-Institut f\"ur Extraterrestrische Physik (MPE), 
National Astronomical Observatories of China, New Mexico State University, 
New York University, University of Notre Dame, 
Observat\'ario Nacional / MCTI, The Ohio State University, 
Pennsylvania State University, Shanghai Astronomical Observatory, 
United Kingdom Participation Group,
Universidad Nacional Aut\'onoma de M\'exico, University of Arizona, 
University of Colorado Boulder, University of Oxford, University of Portsmouth, 
University of Utah, University of Virginia, University of Washington, University of Wisconsin, 
Vanderbilt University, and Yale University.
    
    This research made use of Astropy,\footnote{Astropy Python package: \url{http://www.astropy.org}} a community-developed core Python package for Astronomy \citep{AstropyCollaboration.etal.2013a, AstropyCollaboration.etal.2018a}.

%**********************************************************************%
%                                                                      %
%                              References                              %
%                                                                      %
%**********************************************************************%
\bibliography{references}

%***************************************************************%
%                                                               %
%                           Appendix                            %
%                                                               %
%***************************************************************%
\onecolumn
\appendix
\section[]{Sample python code to correct emission line fluxes}
\label{sec:code}

The sample code below (also available in eletronic format at \url{https://doi.org/10.5281/zenodo.3378041}; \citealp{deAmorim.ValeAsari.2019a}) shows how to implement the line flux correction in Python, and use it to calculate the corrected \Nii/\Ha line ratio. We will assume the data come from single-aperture observations.

First, we define the error function, as given by Eq. \ref{eq:erf}. We also define dictionaries containing the coefficients for both apertures.

\begin{lstlisting}[language=Python, caption=Initial definitions.]
import numpy as np
import scipy.special

def erf_eq1(x, x0, sx):
    a = 0.5
    b = 0.5
    xx = (x - x0) / np.sqrt(2 * sx**2)
    return a * scipy.special.erf(xx) + b

# Single aperture - R/R50 = 0.7
coef_aper = {
    3727: [1.130, 0.1302],
    4861: [1.121, 0.1341],
    5007: [1.131, 0.1262],
    6563: [1.121, 0.1340],
    6584: [1.131, 0.1300],
    6725: [1.130, 0.1316],
    }

# Whole galaxies - R/R50 = 2.0
coef_gal = {
    3727: [1.108, 0.2086],
    4861: [1.078, 0.1944],
    5007: [1.113, 0.1963],
    6563: [1.077, 0.1932],
    6584: [1.098, 0.1944],
    6725: [1.100, 0.1993],
    }
\end{lstlisting}

The user has to provide the emission line flux measurements, as well as \WHa.
For this example, let's generate some random emission line flux and \WHa measurements.

\begin{lstlisting}[language=Python, caption=Generation of random data., firstnumber=29]
# Fix random seed
np.random.seed(42)
    
# Set number of galaxies
Ng = 5

log_WHa_obs = np.random.uniform(1., np.log10(20.), Ng)
F_6563_obs = np.random.rand(Ng)
F_6584_obs = np.random.rand(Ng)
\end{lstlisting}

Here, \WHa is in the $10$ --$20\,\text{\AA}$ range and the fluxes are in the $0$ -- $1$ range (the units are irrelevant in this example). For single-aperture observations, we use the coefficients defined in \texttt{coef\_aper}.
The correction, then, is done as follows. The \texttt{print} statements show the observed and corrected \Nii/\Ha line ratios.

\begin{lstlisting}[language=Python, caption=Flux correction., firstnumber=38]
# Correct line fluxes
F_6563_SFc = F_6563_obs * erf_eq1(log_WHa_obs, *coef_aper[6563])
F_6584_SFc = F_6584_obs * erf_eq1(log_WHa_obs, *coef_aper[6584])

# Check the N2 index, for instance
print('log_WHa_obs =', log_WHa_obs)
print('N2_obs = ', F_6584_obs/F_6563_obs)
print('N2_SFc = ', F_6584_SFc/F_6563_SFc)
\end{lstlisting}

\bsp	% typesetting comment

\label{lastpage}

\end{document}